\documentclass[aps,floats]{revtex4}
\usepackage{amsmath,amssymb}
\usepackage{graphicx,epsfig}

\begin{document}
\bibliographystyle {plain}

\def\oppropto{\mathop{\propto}} 
\def\opsimeq{\mathop{\simeq}}
\def\opoverderline{\mathop{\overline}}
\def\operarrow{\mathop{\longrightarrow}}
\def\opsim{\mathop{\sim}}

\def\fig#1#2{\includegraphics[height=#1]{#2}}
\def\figx#1#2{\includegraphics[width=#1]{#2}}


\title{ Matching between typical fluctuations and  
large deviations in disordered systems : \\
application to the statistics of the ground state energy in 
the SK spin-glass model } 


 \author{ C\'ecile Monthus and Thomas Garel }
  \affiliation{ Institut de Physique Th\'{e}orique, CNRS and CEA Saclay,
 91191 Gif-sur-Yvette, France}

\begin{abstract}
For the statistics of global observables in disordered systems, we discuss the matching between typical fluctuations and large deviations. We focus on the statistics of the ground state energy $E_0$ in two types of disordered models : (i) for the directed polymer of length $N$ in a two-dimensional medium, where many exact results exist (ii) for the Sherrington-Kirkpatrick spin-glass model of $N$ spins, where various possibilities have been proposed. Here we stress that, besides the behavior of the disorder-average $E_0^{av}(N)$ and of the standard deviation $ \Delta E_0(N) \sim N^{\omega_f}$ that defines the fluctuation exponent $\omega_f$, it is very instructive to study the full probability distribution $\Pi(u)$ of the rescaled variable $u= \frac{E_0(N)-E_0^{av}(N) }{\Delta E_0(N)}$ : (a) numerically, the convergence towards $\Pi(u)$ is usually very rapid, so that data on rather small sizes but with high statistics allow to measure the two tails exponents $\eta_{\pm}$ defined as $\ln \Pi(u \to \pm \infty) \sim - \vert u \vert^{\eta_{\pm}}$. In the generic case $1< \eta_{\pm} < +\infty$, this leads to explicit non-trivial terms in the asymptotic behaviors of the moments $\overline{Z_N^n}$ of the partition function when the combination $[\vert n \vert N^{\omega_f}]$ becomes large (b) simple rare events arguments can usually be found to obtain explicit relations between $\eta_{\pm}$ and $\omega_f$. These rare events usually correspond to 'anomalous' large deviation properties of the generalized form $R( w_{\pm} = \frac{E_0(N)-E_0^{av}(N)}{N^{\kappa_{\pm}}} ) \sim e^{- N^{\rho_{\pm}} {\cal R}_{\pm}(w_{\pm})}$ (the 'usual' large deviations formalism corresponds to $\kappa_{\pm}=1=\rho_{\pm}$).

\end{abstract}

\maketitle

\section{ Introduction }

In the field of disordered systems, the interest has been first on 
self-averaging quantities, like the free-energy per degree of freedom,
or other thermodynamic observables that determine the phase diagram.
 However, it has become clear over the years
that a true understanding of random systems has to include the sample-to-sample
fluctuations of global observables, in particular in disorder-dominated phases
where interesting universal critical exponents show up. 
Besides these typical sample-to-sample
fluctuations, it is natural to characterize also the large deviations properties,
since rare anomalous regions are known to play a major role in various properties
of random systems.
 
Among the various global observables that are interesting, 
the simplest one is probably the ground-state energy $E_0$ of a disordered sample.
Since it is the minimal value among the energies of all possible configurations,
the study of its distribution belongs to the field of
 extreme value statistics.
Whereas the case of independent random variables is well 
classified in three universality classes \cite{Gum_Gal}, the problem for 
the correlated energies within a disordered sample remains open
and has been the subject of many recent studies (see for instance \cite{B_JPB_P}
and references therein). 
For many-body models with $N$ degrees of freedom ($N$ spins for disordered spin models,
$N$ monomers for disordered polymers models), the interest lies   

(i) in the scaling behavior of the average
 $E_0^{av}(N)$ and the standard deviation $ \Delta E_0(N)$ 
 with $N$.  Following the definitions of Ref. \cite{Bou_Krz_Mar},
the `shift exponent' $\omega_s$ governs the correction to extensivity
of the averaged value
\begin{eqnarray}
E_0^{av}(N) \simeq N e_0+ N^{\omega_s} e_1 +...
\label{e0av}
\end{eqnarray}
whereas the `fluctuation exponent' $\omega_f $  governs the growth
 of the standard deviation
 \begin{eqnarray}
\Delta E_0(N)  \sim  N^{\omega_f} 
\label{deltae0}
\end{eqnarray}

(ii) in the asymptotic distribution $\Pi (u)$ 
of the rescaled variable
\begin{eqnarray}
u \equiv \frac{E_0 -E_0^{av}(N)}{\Delta E_0(N)}
\label{defu}
\end{eqnarray}
in the limit $N \to \infty$
\begin{equation}
{\cal P}_N(E_0)   \opsimeq_{N \to \infty}  \frac{1}{ \Delta E_0(N)} \  
\Pi \left( u= \frac{
E_0 -E_0^{av}(N)}{ \Delta E_0(N) }  \right) 
\label{scalinge0}
\end{equation}
This scaling function $\Pi(u)$
describes the typical events where the variable $u$ is finite.

(iii) in the large deviations properties. 
In the standard 'large deviation formalism' (see for instance the recent review
\cite{touchette} and references therein), 
one is interested in
the exponentially rare events giving rise to a {\it finite difference} $v$
between the intensive observable $(E_0/N)$ and its averaged value $E_0^{av}(N)/N$
\begin{equation}
 v\equiv \frac{E_0-E_0^{av}(N)}{N}
\label{defv}
\end{equation}
In disordered systems, the probability distributions of these rare events is not necessarily exponentially
small in $N$ but can sometimes involve other exponents $\gamma_{\pm}$
(see examples below in the text)
\begin{eqnarray}
D^-_N( v_- \equiv \frac{E_0-E_0^{av}(N)}{N} <0) 
 && \opsimeq_{N \to \infty} e^{ -N^{\gamma_-} {\cal D}_- (v_-)} \nonumber \\
D^+_N( v_+ \equiv \frac{E_0-E_0^{av}(N)}{N} >0) 
 && \opsimeq_{N \to \infty} e^{ -N^{\gamma_+} {\cal D}_+ (v_+)}
\label{largedeve0}
\end{eqnarray}

In this paper, we discuss these properties
for two types of disordered models : 
 for the directed polymer of length $N$ in a two-dimensional medium, 
where many exact results exist, and 
 for the Sherrington-Kirkpatrick (SK) spin-glass model of $N$ spins, where 
various possibilities have been proposed
 from numerical results or theoretical arguments.
The main conclusions we draw from these two cases are the following :

(a) it is very instructive to study {\it the tails} of
 the full probability distribution $\Pi(u)$ of Eq. \ref{scalinge0} :
 these tails are usually described
by the following form
\begin{eqnarray}
\ln \Pi( u ) && \oppropto_{u \to -\infty} -  (-u)^{\eta_-} +...\nonumber \\
\ln \Pi(u ) && \oppropto_{u \to +\infty} -  u^{\eta_+} +...
\label{defetamu}
\end{eqnarray}
where the two tails exponents $\eta_{\pm}$ are usually different
and in the range $1 \leq \eta_{\pm} \leq +\infty$.
In particular, the very common fits based on generalized Gumbel distributions
are very restrictive and very misleading since they correspond to the unique 
values $\eta_-=1$ and $\eta_+=+\infty$.
We also discuss the consequences of Eq. \ref{defetamu} for the moments 
$\overline{Z_N^n}$ of order $n$
(either positive or negative) of the partition function $Z_N$ at very low temperature.

(b) simple rare events arguments can usually be found to obtain explicit relations
between $\eta_{\pm}$ and $\omega_f$. The probability distributions of these rare events usually correspond
to 'anomalous' large deviation properties of the generalized forms
\begin{eqnarray}
R^-_N( w_- \equiv \frac{E_0-E_0^{av}(N)}{N^{\kappa_-}} <0) 
 && \opsimeq_{N \to \infty} e^{ -N^{\rho_-} {\cal R}_- (w_-)} \nonumber \\
R^+_N( w_+ \equiv \frac{E_0-E_0^{av}(N)}{N^{\kappa_+}} >0) 
 && \opsimeq_{N \to \infty} e^{ -N^{\rho_+} {\cal R}_+ (w_+)}
\label{genelargedeve0}
\end{eqnarray}

The paper is organized as follows.
In Section \ref{sec_dp}, we recall the exact results concerning
the directed polymer in a two-dimensional random medium,
and discuss their meaning for the above points (a) and (b).
In Section \ref{sec_sk}, we discuss the case of the 
Sherrington-Kirkpatrick spin-glass model, and we present numerical results
obtained for small sizes but with high statistics.
Our conclusions are summarized in section \ref{sec_conclusion}.

\section{ Reminder on the directed polymer in a two-dimensional random medium }

\label{sec_dp}

\subsection{ Brief summary of exact results  }

The directed polymer model in a two-dimensional random medium 
(see the review \cite{Hal_Zha})
 is an exactly soluble model that has the following properties : 

(i) a single exponent \cite{Hus_Hen_Fis,Kar,Joh,Pra_Spo}
\begin{eqnarray}
\omega_s=\omega_f=\frac{1}{3} 
\label{omegaDP}
\end{eqnarray}
governs both the correction to extensivity
of the average $E_0^{av}(N)$ (Eq. \ref{e0av})
and the width $\Delta E_0(N)$ (Eq. \ref{deltae0}).

(ii) the rescaled distribution $\Pi(u)$ of Eq. \ref{scalinge0}
 is the Tracy-Widom distribution
of the largest eigenvalue of random matrices ensembles
 \cite{Joh,Pra_Spo,prae}. In particular, the two tails exponents 
of Eq. \ref{defetamu} read
\begin{eqnarray}
\eta_- && = \frac{3}{2} \nonumber \\
\eta_+ && = 3
\label{etapmDP}
\end{eqnarray}

(iii) the exponents of the large deviations forms of Eq. \ref{largedeve0}
are respectively  \cite{derrida_leb,dean_maj,maj_verg,diamondtails}
\begin{eqnarray}
\gamma_- && = 1 \nonumber \\
\gamma_+ && = 2
\label{gammapmDP}
\end{eqnarray}

After this brief reminder of known results, we now turn
to their physical interpretation.

\subsection{ Physical interpretation of the large deviation exponents
in terms of simple rare events   }

As explained in detail in \cite{diamondtails},
the large deviation exponents of Eq. \ref{etapmDP} can be understood as follows

(-) to obtain a ground state energy which
 is extensively lower than the typical,
it is sufficient to draw $N$ anomalously good on-site energies along the ground state path. This will happen with a probability
 $e^{-(cst)N}$ corresponding to
 $\gamma_-= 1 $ of Eq. \ref{gammapmDP}.

(+) to obtain a ground state energy which is extensively higher
 than the typical, one needs to draw $N^2$ bad on-site energies
(i.e. in the whole sample). 
 This will happen with a probability $e^{-(cst)N^2}$ corresponding to
 $\gamma_+= 2 $  of Eq. \ref{gammapmDP}.

Note that in the Asymmetric Exclusion process language, 
the interpretation is that to slow down the traffic, it is sufficient to slow down a single particle, whereas to speed up the traffic, one needs to speed up all particles \cite{derrida_leb}. In the random matrix language, the interpretation is that 
to push the maximal eigenvalue inside the Wigner sea, one needs to reorganize everything,
whereas to pull the maximal eigenvalue outside the Wigner sea, one may leave the
Wigner sea unchanged for the other eigenvalues \cite{dean_maj,maj_verg}.

The fact that these large deviation exponents $\gamma_{\pm}$ can be guessed
via simple physical arguments is an important lesson that is very useful
in other disordered models which are not exactly solvable :
in particular, these arguments can be easily extended to the directed polymer
in a random medium of higher dimensionality \cite{diamondtails},
or to other observables in various models 
\cite{diamondtails,conjugate,firstpassage}.

\subsection{ Matching between typical fluctuation and large deviations   }

For an arbitrary probability distribution,
the typical fluctuations in the bulk and 
the rare fluctuations in the far tails are in general different questions.
However, for the probability distribution of the ground state energy $E_0(N)$ 
(or more generally the probability distributions of other global observables) 
in disordered statistical physics models, 
it is very natural, from a physical point of view, to expect 
some matching between the typical fluctuation scaling regime 
where $E_0-E_0^{av} \sim N^{\omega_f}$
and the large deviations scaling regime 
where $E_0 -E_0^{av} \sim N$.
More precisely, the tails in the regions $u \to \pm \infty$
 of the rescaled distribution $\Pi(u)$ of typical fluctuations should match
smoothly the large deviation region regime
where the variable $v$ of Eq. \ref{defv} is finite,
which corresponds to the
regime where the variable $u$ of Eq. \ref{defu} is of order 
$u \propto N^{1-\omega_f}$.
 If one plugs this scaling into the asymptotic
form of Eq. \ref{defetamu}, and if one insists that one should then recover
the large deviations exponents of Eq. \ref{largedeve0}, one obtains
the very simple relations between exponents \cite{diamondtails}
\begin{eqnarray}
(1-\omega_f)\eta_{\pm}=\gamma_{\pm}
\label{matchingtyplarge}
\end{eqnarray}
For the directed polymer in a two-dimensional random medium, 
these relations are satisfied by the values quoted in 
Eqs \ref{omegaDP},\ref{etapmDP} and \ref{gammapmDP}.
This smooth matching has also been discussed 
in the equivalent problems  
concerning the current in the asymmetric exclusion process \cite{derrida_leb}
and the largest eigenvalue of Gaussian random matrices \cite{dean_maj,maj_verg}.

This matching property between typical fluctuation and large deviations
is again an important lesson that can be used 
in other disordered models which are not exactly solvable.
These relations have been checked in detail 
for the directed polymer
in dimension $d=2,3$ \cite{DPtails} as well as on hierarchical lattices
\cite{diamondtails}. This matching property has also been used recently
for the distribution of the dynamical barriers \cite{conjugate,firstpassage}.
It is also interesting from a physical point of view,
because the asymmetry $\eta_-<\eta_+$ seen in the distribution
of typical events can be seen as a consequence of the asymmetry
of rare events $\gamma_-<\gamma_+ $.

\subsection{ Consequences for the moments of the partition function   }

\label{moments}

Since a direct calculation of the probability distribution of the ground state
of a disordered model is usually very difficult, analytical calculations
 usually focus on the moments $\overline{Z^n_N}$ of the partition function $Z_N$.
Then one can use two types of arguments to relate the distribution of $Z_N=e^{-\beta F_N}$ to the distribution $P_N(E_0)$ of the ground state energy
$E_0(N)$ : 
(1) at very low temperature $T \to 0$, the partition function will be dominated by the ground state $Z_N (\beta \to +\infty) \simeq e^{-\beta E_0(N)}$
(2) moreover in some models, where the disorder-dominated phase $0 \leq T<T_c$
is governed by a zero-temperature fixed point, one expects that 
in the whole region of temperatures $0 \leq T<T_c$, the probability
distribution of the free-energy $F_N$ will actually have the same properties
as the distribution of $E_0$.
Since (2) is valid for the directed polymer model,
but cannot be taken for granted in all disordered models,
we will restrict here to the point of view (1) of 
very low temperature $T \to 0$.

There exists a simple argument that has been proposed by Zhang \cite{Hal_Zha} 
on the specific case on the directed polymer,
 that relates the scaling behaviors of the 
moments $\overline{Z^n_N}$ with the size $N$ and with the replica index $n$
to the properties of $P_N(E_0)$.
The idea is to evaluate the moments by using the rescaled distribution
of Eq. \ref{scalinge0}
\begin{eqnarray}
\overline{Z_N^n} = \int dE_0 P_N(E_0) e^{- \beta n E_0 }
\simeq  \frac{1}{ \Delta E_0(N)} e^{- \beta n E_0^{av}(N)} \ \int du  
\Pi ( u )  e^{- \beta n \Delta E_0(N) u }
\label{zhang}
\end{eqnarray}
For the case $n>0$ considered by Zhang \cite{Hal_Zha}, 
the integral can be then evaluated by a saddle point method
in the region $u \to - \infty $, where one may use the asymptotic
behavior of Eq. \ref{defetamu} with the exponent $\eta_- >1 $ :
the saddle point is of order
\begin{eqnarray}
u^* \propto (\beta n \Delta E_0(N)  )^{\frac{1}{\eta_- -1}}
\label{usaddle}
\end{eqnarray}
that should be large $u^* \gg 1$, and one obtains
\begin{eqnarray}
\overline{Z_N^{n>0}} \propto 
e^{- \beta n E_0^{av}(N) + (cst) (\beta n \Delta E_0(N) 
 )^{\frac{\eta_-}{\eta_- -1} } }
\label{zhang2}
\end{eqnarray}
For the case $n<0$, the equivalent calculation yields in term
of the other tail exponent $\eta_+ >1 $ :
\begin{eqnarray}
\overline{Z_N^{n<0}} \propto 
e^{- \beta n E_0^{av}(N)+ (cst) ( \beta (-n) \Delta E_0(N)  )^{\frac{\eta_+}
{\eta_+ -1} } }
\label{zhang3}
\end{eqnarray}

For the directed polymer in a  two-dimensional random medium, 
one obtains, using $\Delta E_0(N) \propto N^{\omega_f}$ 
with the explicit values of 
Eqs \ref{omegaDP},\ref{etapmDP}
\begin{eqnarray}
\overline{Z_N^{n>0}} \propto 
e^{- \beta n E_0^{av}(N) + (cst) (\beta n)^3 N  }
\label{zhangnposdp}
\end{eqnarray}
where one recognizes the combination $(n^3 N)$ that appears
in the Bethe Ansatz replica calculation of Ref. \cite{Kar}.
Moreover, in Zhang's argument \cite{Hal_Zha},
one actually imposes that the non-trivial term of Eq. \ref{zhang2} 
should be extensive in $N$ 
(because for positive integer $n$, the moments of the partition function 
can be formulated in terms of the iteration of some transfer matrix,
and thus they have to diverge
 exponentially in $N$ with some Lyapunov exponent)
to obtain the relation $\omega_f \frac{\eta_-}{\eta_- -1)} =1$
(which is equivalent here to the relation of Eq. \ref{matchingtyplarge}
obtained previously by the rare event interpretation).

For $n<0$, the obtained behavior
\begin{eqnarray}
\overline{Z_N^{n<0}} \propto 
e^{- \beta n E_0^{av}(N) + (cst)  (\beta (-n))^{\frac{3}{2}} N^{\frac{1}{2}}   }
\label{zhangnnegdp}
\end{eqnarray}
is rather different : the only extensive contribution of order $N$ in the exponential
comes from $E_0^{av}(N)$.
The leading contribution due to fluctuations is only of subleading order $ N^{\frac{1}{2}}$,
and it involves a non-integer power of the replica index $(-n)$.
To the best of our knowledge, the behavior of these negative moments
has not been much discussed in the literature, in contrast to 
the case $n>0$.

These saddle-point calculations based on the facts that the tails
exponents satisfy $\eta_{\pm}>1$ can be very useful in other non-exactly
soluble models, for instance in the Sherrington-Kirkpatrick spin-glass model
that we now consider.

\section{ Sherrington-Kirkpatrick spin-glass model }

\label{sec_sk}

For short-ranged spin-glasses in any finite dimension $d$,
it has been proven that the fluctuation
exponent of Eq. \ref{deltae0} is exactly $\theta_f=d/2$ \cite{We_Ai}.
Accordingly, the rescaled distribution ${\Pi }(u)$ of Eq. (\ref{scalinge0}) 
was numerically found to be Gaussian in $d=2$ and $d=3$ 
\cite{Bou_Krz_Mar},
suggesting some Central Limit theorem.
On the contrary, in mean-field spin-glasses, the width
does not grows as $N^{1/2}$ and the distribution is not Gaussian,
as will be discussed in more details in this section.
Studies on long-ranged one-dimensional spin-glasses \cite{KKLJH}
have confirmed that non-mean-field models are characterized by Gaussian
distributions, whereas mean-field models are not.

\subsection{ Brief summary of previous works}

The statistics of the ground state energy of the Sherrington-Kirkpatrick spin-glass model \cite{SKmodel}
\begin{eqnarray}
 H = - \sum_{1 \leq i <j \leq N} J_{ij} S_i S_j
\label{defSK}
\end{eqnarray}
where the couplings $J_{ij}$ are random quenched variables
of zero mean $\overline{J}=0$ and of variance $\overline{J^2}=1/N $,
 has been much studied recently
 with the following outputs :

(i) there seems to be a consensus 
(see for instance \cite{palassini,Bou_Krz_Mar,
boettcher09} and references therein) on the shift exponent of Eq. \ref{e0av}
\begin{eqnarray}
\omega_s = \frac{1}{3}
\label{shiftSK}
\end{eqnarray}
whereas the `fluctuation exponent' $\omega_f $ is still under debate 
between the value (see \cite{aspelmeier_MY,Bou_Krz_Mar,boettcher09}
and references therein)
 \begin{eqnarray}
\omega_f = \frac{1}{4} 
\label{1/4}
\end{eqnarray}
and the value (see \cite{crisanti,aspelmeier_BMM,palassini,rizzo} and references therein)
 \begin{eqnarray}
\omega_f = \frac{1}{6} 
\label{1/6}
\end{eqnarray}

(ii)  the asymptotic distribution $\Pi(u)$ of Eq. \ref{scalinge0}
has been measured numerically by various authors
(see \cite{palassini,KKH,boettcher05} and references therein),
 but unfortunately it has almost always been
fitted by 'generalized Gumbel distributions' of the form $e^{m(u-e^{u}) }$
 containing a single free-parameter $m$ for the shape.
However these fits are very restrictive and very misleading
since the tails exponents are fixed to be 
\begin{eqnarray}
{\rm Generalized Gumbel : } \ \ \eta_- && =1 \nonumber \\
 \ \eta_+ && =+\infty
\label{geneGumbel}
\end{eqnarray}
for any value of the parameter $m$.
In this paper, we propose instead that these 
exponents are in the range $1<\eta_{\pm}<+\infty$.

(iii)  the large deviation properties have been also very controversial.
In \cite{andreanov}, numerical results have been interpreted
with the following values for the exponents $\gamma_{\pm}$ of Eq. 
\ref{largedeve0}
\begin{eqnarray}
{\rm Ref. [29] : } \ \ \gamma_- && \simeq 1.2 \nonumber \\
  \gamma_+ && \simeq 1.5
\label{ldandeanov}
\end{eqnarray}
Other proposals are (see \cite{rizzo} and references therein)
\begin{eqnarray}
{\rm Ref. [26] : } \ \ \gamma_- && =1 \nonumber \\
 \gamma_+ && =2
\label{ldrizzo}
\end{eqnarray}

After this brief summary of conflicting proposals,
we now turn to the analysis
along the same line as in the previous section concerning
the directed polymer model.

\subsection{ Discussion of simple rare events   }

The simplest rare events one may consider for the SK model
are the following :

(-) to obtain a ground state energy which
 is much lower than the typical,
it is natural to consider the anomalous ferromagnetic samples \cite{andreanov}
that appear with a small probability of order $e^{-(cst)N^2}$
(one needs to draw $N^2$ positive couplings in Eq. \ref{defSK}),
and that will corresponds to anomalously low energy of order 
$E_0 \propto -(cst)N^{3/2}$. These events corresponds to the 'very large deviation' of the generalized form of Eq. \ref{genelargedeve0} with the values \cite{andreanov}
\begin{eqnarray}
{\rm Ferro : } \ \ \rho_- && = 2 \nonumber \\
 \ \ \ \ \kappa_- && = \frac{3}{2}
\label{ferro}
\end{eqnarray}
This form has been checked numerically in \cite{andreanov}.

(+) to obtain a ground state energy which is much higher
 than the typical, one could consider the anomalous antiferromagnetic samples
that appear with a small probability of order $e^{-(cst)N^2}$
(one needs to draw $N^2$ negative couplings) and that will
give an energy extensively higher. In the large deviation form of 
Eq. \ref{largedeve0}, this would corresponds to
\begin{eqnarray}
{\rm Antiferro : } \ \ \gamma_+ = 2 
\label{antiferro}
\end{eqnarray}
This value corresponds to the proposal of Eq. \ref{ldrizzo} 
from Ref. \cite{rizzo}, but disagrees with the numerical proposal 
of Eq. \ref{ldandeanov} from Ref. \cite{andreanov}. The question is whether
to obtain an extensively higher energy, it is sufficient to draw
anomalously only a number of order $N^{1.5}$ random couplings instead of $N^2$.
We are presently not aware of any simple argument
 in favor of this smaller power $N^{1.5}$.

\subsection{ Matching between typical fluctuation and large deviations   }

In the (+) region, the matching
between typical fluctuation and rare events leads to the same relation
as in Eq. \ref{matchingtyplarge}
\begin{eqnarray}
(1-\omega_f)\eta_{+}=\gamma_{+}
\label{matchingSKplus}
\end{eqnarray}
In particular, the possible values of $\gamma_+$ and $\omega_f$
leads to the following values for the tail exponent $\eta_+$ :
\begin{eqnarray}
{\rm If } \ \ \gamma_+ = 2 \ \  : 
\eta_+^{\left(\omega_f = \frac{1}{4}  \right)} &&=
\frac{8}{3}=2.6666.. \nonumber \\
\eta_+^{\left(\omega_f = \frac{1}{6}  \right)} &&=
\frac{12}{5}=2.4
\label{etaplusantiferro}
\end{eqnarray}
or
\begin{eqnarray}
{\rm If } \ \ \gamma_+ = \frac{3}{2} \ \  : 
\eta_+^{ \left(\omega_f = \frac{1}{4}  \right)} && =
2  \nonumber \\ 
\eta_+^{\left(\omega_f = \frac{1}{6}  \right)} && =
\frac{9}{5}=1.8
\label{etaplus3/2}
\end{eqnarray}

In the (-) region, the matching
between typical fluctuation and the very large deviations
of Eq. \ref{ldandeanov} leads to the relation
\begin{eqnarray}
\left(\kappa_- -\omega_f \right)\eta_{-}=\rho_-
\label{matchingSKmoinsverylarge}
\end{eqnarray}
Using the values of Eq. \ref{ferro}
one obtains the two possible values for $\eta_-$
\begin{eqnarray}
{ \rm If  } \ \ \  \omega_f=\frac{1}{4} \ \ : \eta_- &&=
 \frac{8}{5}=1.6  \nonumber \\ 
{ \rm If  } \ \ \  \omega_f=\frac{1}{6} \ \ \eta_-&&=
\frac{3}{2}=1.5
\label{etamoins}
\end{eqnarray}

If this matching works, the region of large deviation of Eq. 
\ref{largedeve0} which is between the typical region
and the very large deviation region, is constrained by consistency
to involve the exponent
\begin{eqnarray}
\gamma_-=\left(1 -\omega_f \right)\eta_{-}
\label{matchingSKmoins}
\end{eqnarray}
The two possible values read
\begin{eqnarray}
{ \rm If  } \ \ \  \omega_f=\frac{1}{4} \ \ :\gamma_- &&=
 \frac{6}{5}=1.2 \nonumber \\  
{ \rm If  } \ \ \  \omega_f=\frac{1}{6} \ \ : \gamma_- && =
\frac{5}{4}=1.25
\label{gammamoins}
\end{eqnarray}
Both are close to the numerical value of Eq. \ref{ldandeanov}
proposed by Ref. \cite{andreanov}.
Both disagree with the value $\gamma_-=1$ of Eq. \ref{ldrizzo}
used in replica calculations of \cite{rizzo}.

\subsection{ Consequences for the moments of the partition function   }

As explained in detail in section \ref{moments},
the moments of the partition function $Z_N$
are then expected to follow Eqs \ref{zhang2} and \ref{zhang3}
\begin{eqnarray}
\overline{Z_N^{n>0}} && \propto 
e^{- \beta n E_0^{av}(N) + (cst) (\beta n N^{\omega_f} 
 )^{\frac{\eta_-}{\eta_- -1} } }
\nonumber \\
\overline{Z_N^{n<0}} && \propto 
e^{- \beta n E_0^{av}(N) + (cst) ( \beta (-n) N^{\omega_f}   )^{\frac{\eta_+} 
{\eta_+ -1} } }
\label{zhangSK}
\end{eqnarray}

For positive $n>0$ :
 the two possible values of $\omega_f$ and 
of the associated tail exponent $\eta_-$ 
(see Eq. \ref{etamoins}) correspond to the behaviors
\begin{eqnarray}
{ \rm If  } \ \ \  \omega_f=\frac{1}{4} \ \ : \overline{Z_N^{n>0}} && \propto 
e^{- \beta n E_0^{av}(N) + (cst) (\beta n N^{\frac{1}{4}} 
 )^{\frac{8}{3} } }
\nonumber \\
{ \rm If  } \ \ \  \omega_f=\frac{1}{6} \ \ : \overline{Z_N^{n>0}} && \propto 
e^{- \beta n E_0^{av}(N) + (cst) (\beta n N^{\frac{1}{6}})^{ 3 }  }
\label{zhangSKpos}
\end{eqnarray}
We note that in both cases, the non-trivial part is sub-extensive in $N$,
in contrast to the replica calculations 
of \cite{rizzo}, but in agreement with the replica calculations of \cite{am,cirano}. It is also clear that 
the non-trivial part $(n^3 N^{1/2})$ for the case $\omega_f=\frac{1}{6} $,
is simpler than the term $(n^{8/3} N^{2/3})$
 for the case $\omega_f=\frac{1}{4} $. In both cases, the powers of $n$ that appear are different from the value $n^5$ of perturbative replica calculations
\cite{kondor,rizzo}.

For negative $n<0$ : for the case $\gamma_+ = 2$ of Eq. \ref{etaplusantiferro},
the possible behaviors are 
\begin{eqnarray}
{ \rm If  } \ \ \  \omega_f=\frac{1}{4} \ \ :\overline{Z_N^{n<0}} && \propto 
e^{- \beta n E_0^{av}(N) + 
(cst) ( \beta (-n) N^{\frac{1}{4}} )^{\frac{8}{5} } }
\nonumber \\
{ \rm If  } \ \ \  \omega_f=\frac{1}{6} \ \ :\overline{Z_N^{n<0}} && \propto 
e^{- \beta n E_0^{av}(N) + 
(cst) ( \beta (-n) N^{\frac{1}{6}} )^{\frac{12}{7} } }
\label{zhangSKneg}
\end{eqnarray}
For the case $\gamma_+ = 3/2$ of Eq. \ref{ldandeanov} proposed in Ref. \cite{andreanov},
 the behavior of the moments can be similarly evaluated using Eq. 
\ref{etaplus3/2}.
Again in all cases, the non-trivial part is sub-extensive in $N$,
as already proposed in \cite{dotsenko}.
Concerning the powers of $(-n)$, the exponent $(12/7)$ for the case
$\omega_f=\frac{1}{6}$ is in agreement with the replica calculations of 
\cite{rizzo}.

\subsection{ Numerical results 
for small sizes and large statistics of samples }

\begin{figure}[htbp]
 \includegraphics[height=6cm]{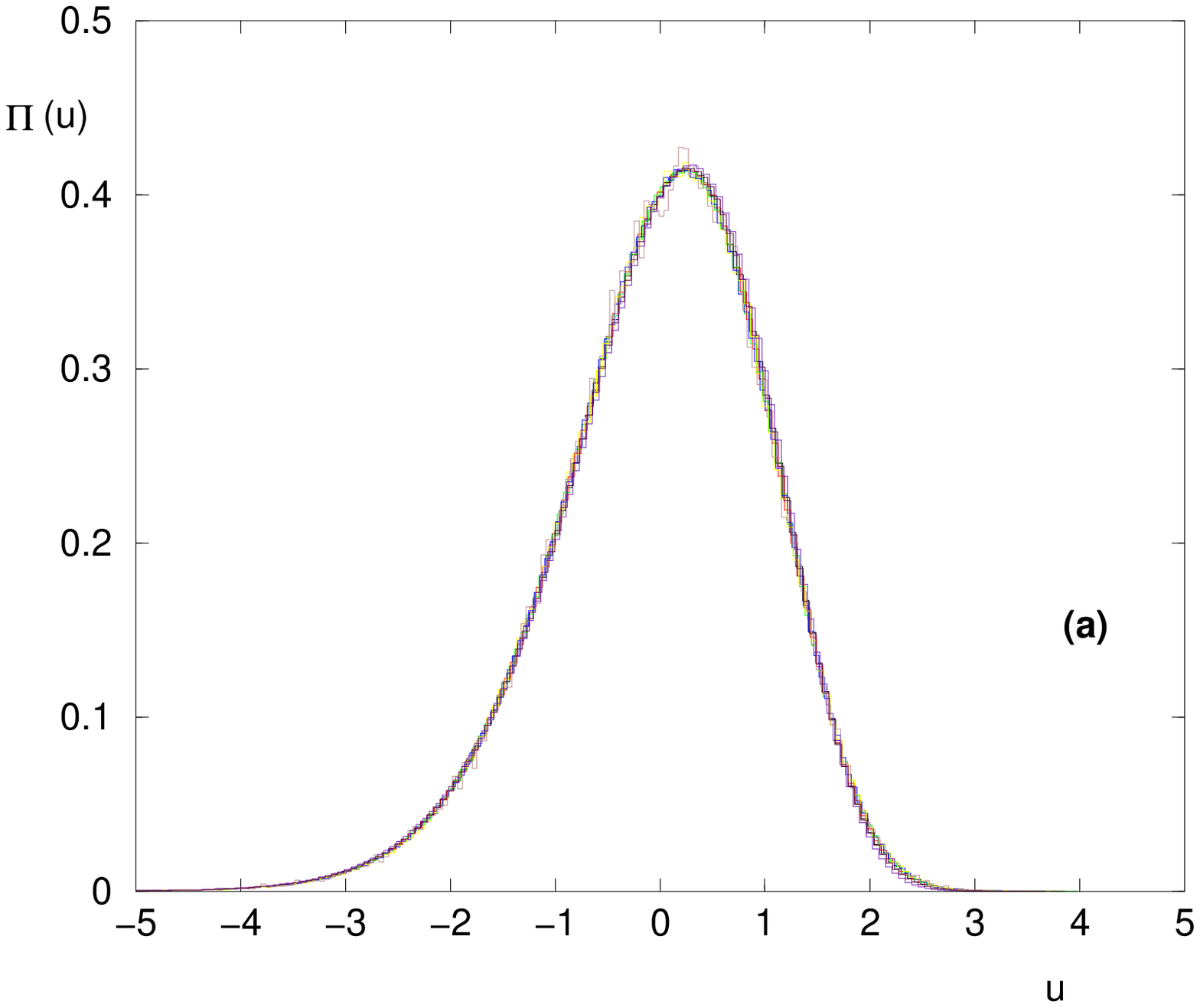}
\hspace{2cm}
 \includegraphics[height=6cm]{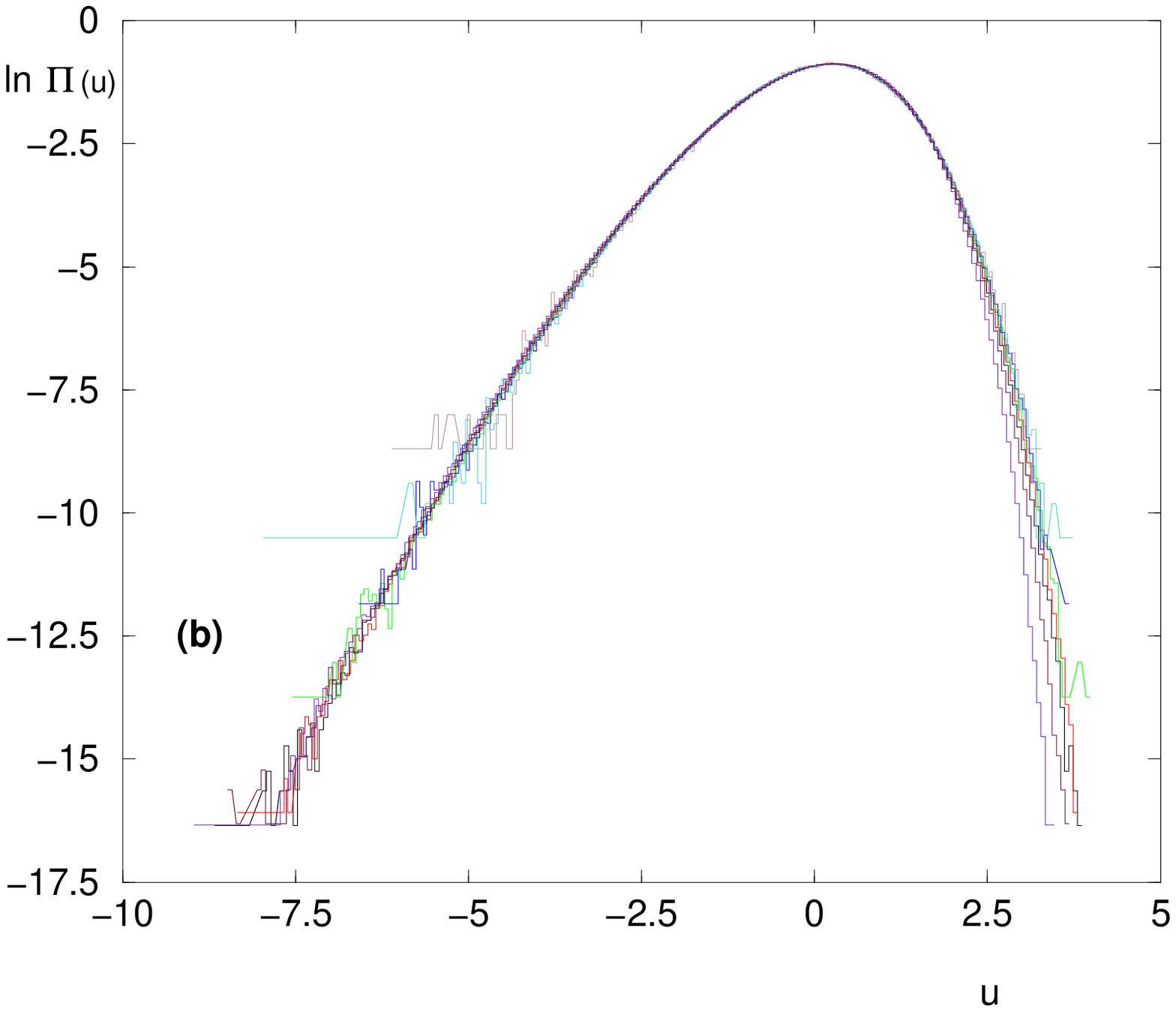}
\vspace{1cm}
\caption{ Rescaled probability distribution of the ground state energy $E_0(N)$
in the SK model :
(a) the histograms $\Pi_N(u)$ 
of the rescaled variable $u$ of Eq. \ref{defu}
measured for even sizes in the range $6 \leq N \leq 20$ almost coincide :
this shows that the convergence in $N$ towards the asymptotic form
is very rapid.
(b) same data in log-scale to see the tails : one sees that the left tail
does not change, whereas finite-size effects are visible on the right tail. }
\label{fighisto}
\end{figure}

\begin{figure}[htbp]
 \includegraphics[height=6cm]{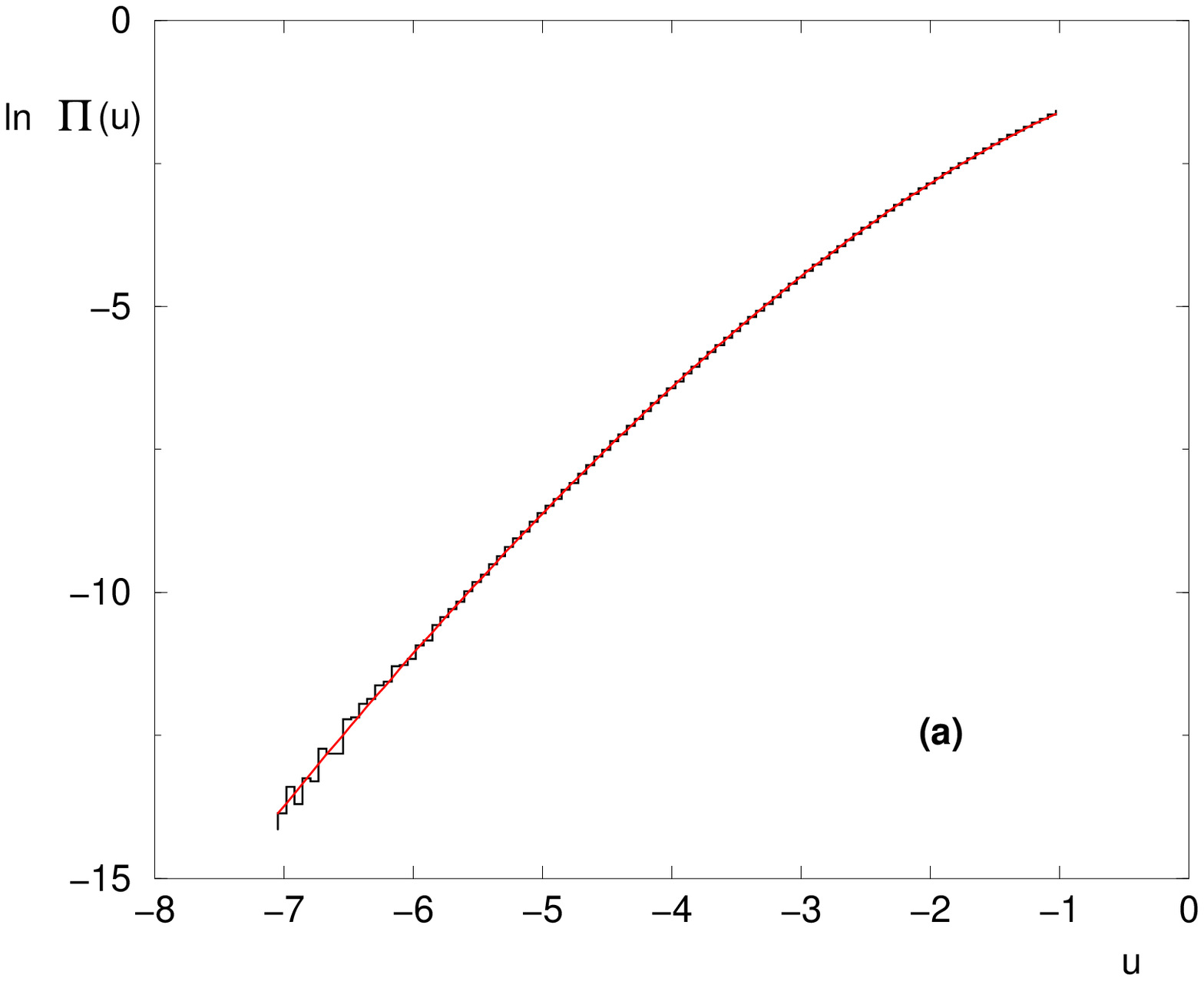}
\hspace{2cm}
 \includegraphics[height=6cm]{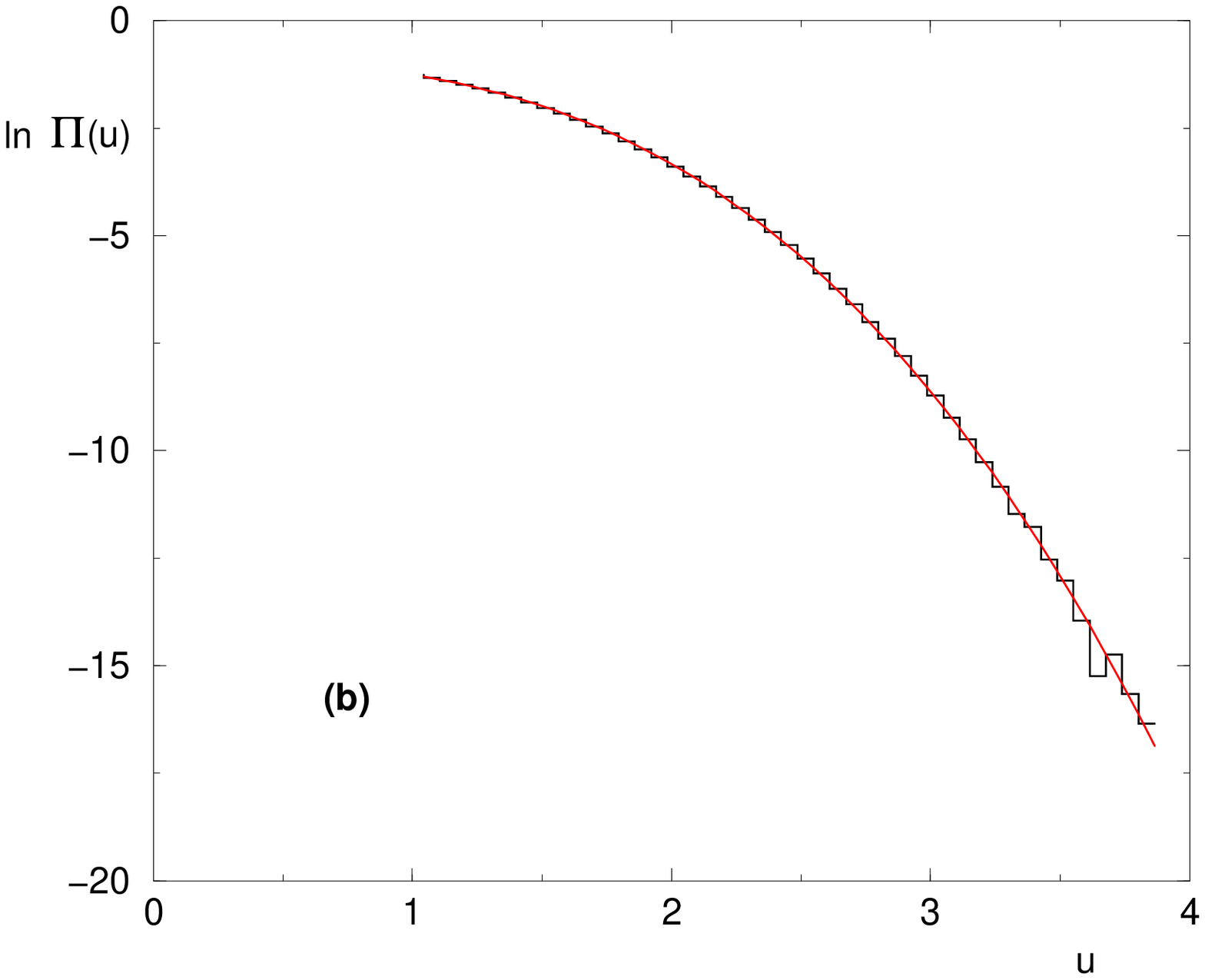}
\vspace{1cm}
\caption{ Tails exponents of the rescaled distribution $\Pi_N(u)$ :
examples of fits of our numerical rescaled 
histogram (step function) corresponding to the size $N=10$
(a) The smooth curve corresponds to the best
 three-parameter fit in the range $u \leq -1$ 
by the form $(a_0 - a_1(-u)^{\eta_-})$ : the left tail exponent is
of order $\eta_- \simeq 1.5$.
(b) The smooth curve corresponds to the best
 three-parameter fit in the range $u \geq 1$ 
by the form $(a_0 - a_1 u^{\eta_+})$ : the right tail exponent is
of order $\eta_+ \simeq 2.8$. }
\label{figfit}
\end{figure}

Most numerical works on the distribution of the ground state energy
in the SK model have followed the strategy to study the biggest sizes $N$
as possible, to measure the averaged value and the variance
(see \cite{palassini,Bou_Krz_Mar,boettcher09} and references therein).
An opposite strategy has been followed in Ref. 
\cite{boettcher05} where an exact enumeration of the
disordered samples with the binomial distribution $\pm J$ was
performed for small sizes. As mentioned in \cite{boettcher09},
the results for the rescaled distribution $\Pi(u)$ at $N=9$ are
already very good when compared to the results for larger $N$.
In other cases, we have also found that the distribution
of rescaled variables converge much more rapidly than other
observables \cite{conjugate,firstpassage,rgmaster}.
In the following, we thus follow the same strategy :
we study the distribution of $E_0(N)$ for the small sizes
with a high statistics of disordered samples.

On Fig. \ref{fighisto} (a), we show the measured histograms $\Pi_N(u)$
of the the rescaled variable $u$ of Eq. \ref{defu}
for even sizes in the range
$6 \leq N \leq 20$ with a statistics of $2.10^8 \geq n_s(N) \geq 13.10^4$
disordered samples : one clearly sees that all these histograms 
almost coincide. Our conclusion is thus that
the convergence in $N$ towards the asymptotic form
is very rapid, so that these small-size data should provide a reliable measure
of the asymptotic $\Pi(u)$.  As explained before, we are mainly interested into
the tails exponents $\eta_{\pm}$ of Eq. \ref{defetamu} :
as shown on Fig. \ref{fighisto} (b) the convergence of the left tail is extremely good, whereas the convergence of the right tail presents much stronger
finite-size effects.

Let us first consider the left tail.
The three-parameter fit of $\Pi(u)$ in the range $u \leq -1$ 
by the form $(a_0 - a_1(-u)^{\eta_-})$ yields the value
(see Fig \ref{figfit} (a))
\begin{eqnarray}
\eta_- \simeq 1.5
\label{etamoinsres}
\end{eqnarray}
that corresponds exactly to the value associated to $\omega_f=1/6$ (see Eq. \ref{etamoins}).
Of course, it is probably not far enough from the alternative value
$\eta_-=1.6$ corresponding to $\omega_f=1/4$ (see Eq. \ref{etamoins})
to really rule out the value $\omega_f=1/4$.

Let us now turn to the right tail. 
The three-parameter fit of $\Pi(u)$ in the range $u \geq 1$ 
by the form $(a_0 - a_1 u^{\eta_+})$ yields values for
$\eta_+ $ that are less precise,
as a consequence of the finite-size effects visible 
on Fig. \ref{fighisto} (b). We have already found in other studies
that the right tail is usually more difficult to measure than the left tail 
\cite{diamondtails}.
Nevertheless our non-precise values of $\eta_+$ in the range $[2.4,2.9]$
(see Fig \ref{figfit} (b))
 seem more compatible with the value $\gamma_+=2$
 than with the value $\gamma_+=1.5$
(see Eq. \ref{etaplusantiferro} and \ref{etaplus3/2}).

\subsection{ Final discussion}

In summary, even if a definitive agreement on the precise value
 of the fluctuation exponent $\omega_f$ remains difficult to reach
(see \cite{palassini,Bou_Krz_Mar,boettcher09} and references therein),
our conclusions concerning the SK model are the following :

(i) the numerical measure of the left tail exponent $\eta_-$
is in agreement with the matching argument
 based on rare ferromagnetic samples 
described by the very-large deviation
 form of Eq. \ref{genelargedeve0} with the values of Eq. \ref{ferro} from 
Ref. \cite{andreanov}.
Then the large deviation form of Eq. \ref{largedeve0} is constrained to involve an exponent 
$\gamma_- $ given by Eq. \ref{matchingSKmoins}
\begin{eqnarray}
\gamma_-= (1-\omega_f) \eta_- = \frac{2 (1-\omega_f)}{(\frac{3}{2}-\omega_f)}
\label{gammamoinsfinal}
\end{eqnarray}
This explains the numerical value of Eq. \ref{ldandeanov}
proposed in Ref. \cite{andreanov}, and excludes 
 the value $\gamma_-=1$ of Eq. \ref{ldrizzo}
used in the replica calculations of \cite{rizzo}.
We note moreover that this 'usual large deviation value' $\gamma_-=1$
would be satisfied only for the value $\omega_f=1/3$, i.e. only if the fluctuation exponent 
$\omega_f$ would coincide with the shift exponent $\omega_s=1/3$ of Eq. \ref{shiftSK}.

(ii) although less precise, the numerical measure of the right tail exponent $\eta_+$ is more in favor of the large deviation exponent $\gamma_+=2$, that can be justified with a simple rare events argument (see Eq. \ref{antiferro}).

(iii) finally, the facts that the tails exponents satisfy $\eta_{\pm} >1$ induces non-trivial behavior for the moments of the partition function (see Eqs. \ref{zhangSK})
when $(\vert n \vert N^{\omega_f})$ becomes large.
In particular, from Eqs \ref{zhangSKpos} and \ref{zhangSKneg}, our conclusion is that
the only extensive term in $N$ comes from the trivial term $E_0^{av}(N)$ both for negative and positive $n$. Moreover, the non-trivial sub-extensive terms can a priori involve non-integer powers of the replica index $n$.

\section{Conclusion}

\label{sec_conclusion}

In this paper, we have discussed the statistics of the ground state energy $E_0(N)$ in two types of disordered models: (i) for the directed polymer of length $N$ in a two-dimensional medium, where many exact results exist (ii) for the Sherrington-Kirkpatrick spin-glass model, where  various possibilities are still under debate both numerically and theoretically. 
Our main conclusions are the following. Besides the behavior of the disorder-average $E_0^{av}(N)$ and of the standard deviation $ \Delta E_0(N) \sim N^{\omega_f}$, it is very instructive to study the full probability distribution $\Pi(u)$ of the rescaled variable $u= \frac{E_0(N)-E_0^{av}(N) }{\Delta E_0(N)}$ : 

(a) numerically, the convergence towards $\Pi(u)$ is usually very rapid, so that data on rather small sizes but with high statistics allow to measure the tails exponents $\eta_{\pm}$ that satisfy generically $1< \eta_{\pm} < +\infty$
(whereas the very common fits based on generalized Gumbel distributions
correspond to the unique 
values $\eta_-=1$ and $\eta_+=+\infty$). Moreover, if one wishes to measure tails
beyond the region probed via simple sampling, one may uses a Monte-Carlo procedure
in the disorder, as done in \cite{KKH} for the SK model,
and in \cite{DPtails} for the directed polymer model.

 (b) simple rare events arguments can usually be found to obtain explicit relations between $\eta_{\pm}$ and $\omega_f$. These rare events usually correspond to 'anomalous' large deviation properties of the generalized form $R_N( w_{\pm} = \frac{E_0(N)-E_0^{av}(N)}{N^{\kappa_{\pm}}} ) \sim e^{- N^{\rho_{\pm}} {\cal R}_{\pm}(w_{\pm})}$ (the 'usual'' large deviations formalism corresponding to $\kappa_{\pm}=1=\rho_{\pm}$ is too restrictive for disordered models,
as shown on explicit examples in the text).

(c) We have also discussed the consequences of $1< \eta_{\pm} < +\infty$
 for the moments $\overline{Z_N^n}$ of order $n$
(either positive or negative) of the partition function $Z_N$.
In the regime where $[\vert n \vert N^{\omega_f}]$ becomes large, a saddle-point calculation leads to explicit non-trivial terms in the asymptotic behaviors of the moments $\overline{Z_N^n}$ of the partition function.

We have shown in detail how this analysis for the directed polymer is 
in agreement with all known exact results. For the SK model, we have explained
how this analysis agrees or disagrees
 with various possibilities debated in the literature.

In conclusion, we believe that this type of analysis based on the matching
 between typical fluctuations and rare events is very useful 
to study disordered systems. Here we have focused on the statistics of
the ground state energy, but it can also be used for other global observables
such as the maximal dynamical barrier of a disordered sample \cite{conjugate,firstpassage}, 
or for the statistics of large excitations in ferromagnets and spin-glasses
\cite{diamondtails}.

\end{document}